\documentclass[10pt]{wlscirep}
\usepackage[utf8]{inputenc}
\usepackage[T1]{fontenc}
\usepackage{placeins}
\usepackage{setspace}
\usepackage{soul}
\newcommand{\changed}[1]{\textcolor{black}{#1}}

\usepackage{graphicx} 
\usepackage[inkscapelatex=false]{svg}
\usepackage{pdfpages}
\usepackage{tikz}
\usepackage{amsthm}
\usepackage{amsmath}
\usepackage{amssymb}
\usepackage{placeins}
\usepackage{xcolor}

\usepackage{caption}
\usepackage{subcaption}
\usepackage{algorithm}
\usepackage{algpseudocode}
\usepackage{multirow}
\usepackage[flushleft]{threeparttable}
\usepackage{bm}

\NewDocumentCommand{\grad}{e{_^}}{
  \mathop{}\!%
  \nabla
  \IfValueT{#1tli}{_{\!#1}}%
  \IfValueT{#2}{^{#2}}%
}

\title{Forward-only learning in memristor arrays \\ with month-scale stability}

\author[1]{Adrien Renaudineau} \author[2,+]{Mamadou Hawa Diallo} \author[1,+]{Th\'eo Dupuis} \author[2,+]{Bastien Imbert}
 \author[1]{Mohammed Akib Iftakher}  \author[3]{Kamel-Eddine Harabi} \author[1]{Cl\'ement Turck} \author[3]{Tifenn Hirtzlin}   \author[1]{Djohan Bonnet}  
 \author[2]{Franck Melul}  \author[2]{Jorge-Daniel Aguirre-Morales} \author[3]{Elisa Vianello}
\author[2]{Marc Bocquet} \author[2]{Jean-Michel Portal} \author[1,*]{Damien Querlioz} 
\affil[1]{Universit\'e Paris-Saclay, CNRS, Centre de Nanosciences et de Nanotechnologies,  Palaiseau, France}
\affil[2]{Aix-Marseille Université, CNRS, Institut Matériaux Microélectronique Nanosciences de Provence, Marseille, France}
\affil[3]{Universit\'e Grenoble-Alpes, CEA, LETI,   Grenoble, France}
\affil[*]{damien.querlioz@universite-paris-saclay.fr}
\affil[+]{These authors contributed equally to this work}

\begin{abstract}
Turning memristor arrays from efficient inference engines into systems capable of on-chip learning has proved difficult. Weight updates have a high energy cost and cause device wear, analog states drift, and backpropagation requires a backward pass with reversed signal flow. Here we experimentally demonstrate learning on standard filamentary HfO$_x$/Ti arrays that addresses these challenges with two design choices. First, we rely on forward-only training algorithms in the Forward-Forward family that use only inference-style operations. Second, we use sub-1~V reset-only, single-pulse updates that cut energy and yield stable analog states. \changed{We first validate fully in-hardware single-layer learning, where both analog MAC operations and sub-1~V programming updates are performed on chip. We then train two-layer classifiers on an ImageNet-resolution four-class task using a hardware-in-the-loop protocol on arrays up to 8,064 devices: all conductance reads and programming updates are physical, and multilayer MACs are emulated in software from the measured conductances.} Two forward-only variants, two-pass supervised Forward-Forward and a single-pass competitive rule, achieve test accuracies of 89.5\% and 89.6\%, respectively; a reference experiment using backpropagation reaches 90.0\%. Across five independent runs per method, these accuracies are indistinguishable within statistical uncertainty. \changed{Trained models retain accuracy for at least one month under ambient conditions and remain stable under an aggressive 25~h read-disturb protocol, whereas a matched set-based Write\&Verify baseline loses nearly three percentage points of test accuracy.} \changed{At the array level,} sub-1~V reset updates use 460 times less energy than conventional program-and-verify programming and require just 46\% more energy than inference-only operation. Together, these results establish forward-only, sub-1~V learning on standard filamentary stacks at array scale, outlining a practical, pulse-aware route to adaptive edge intelligence.
\end{abstract}

\begin{document}
%TC:ignore
\maketitle

%\linenumbers

\thispagestyle{empty}

%TC:endignore

\section*{Introduction}

Memristor arrays implement analog multiply-accumulate (MAC) operations by Ohm’s and Kirchhoff’s laws, enabling remarkably low-energy inference in dense crossbars \cite{kim20214k,wan2022compute,hung2021four,huang2024memristor,li2020cmos,yousuf2025layer,ambrogio2023analog,le202364}.
As large-scale memristor-based inference has matured, the next frontier is on-chip learning: adapting models directly where data are acquired. This capability is critical for edge scenarios such as medical sensing or predictive maintenance, where models must specialize in situ to nonstationary conditions and potentially confidential information. 
However, turning memristor arrays from efficient inference engines into learning systems has proved difficult because training stresses both devices and architectures. At the architecture level, multilayer training with backpropagation is poorly matched to crossbar datapaths: it requires a backward pass (transpose-array access or reversed signal flow) and storage of intermediate activations, adding circuit complexity and energy overhead relative to inference. At the device level, learning requires frequent and fine-grained weight updates; in filamentary devices these updates are typically implemented with program-and-verify loops \cite{esmanhotto2022experimental,rao2023thousands}, which consume substantial energy, accelerate wear, often require write voltages above 1~V, and must contend with drift of analog conductance states. Nonfilamentary devices can avoid program-and-verify \cite{stecconi2022filamentary,park2024multi}, but they generally still require bit-line voltages above one volt and trade learning capability for retention and stability.  Together, these constraints suggest that practical crossbar learning needs (i)  a learning rule that can be executed with inference-style operations and (ii) a write primitive compatible with many low-energy updates at CMOS-friendly voltages while producing stable analog states.

\begin{figure}[h]
    \centering
    \includesvg[width=1\linewidth]{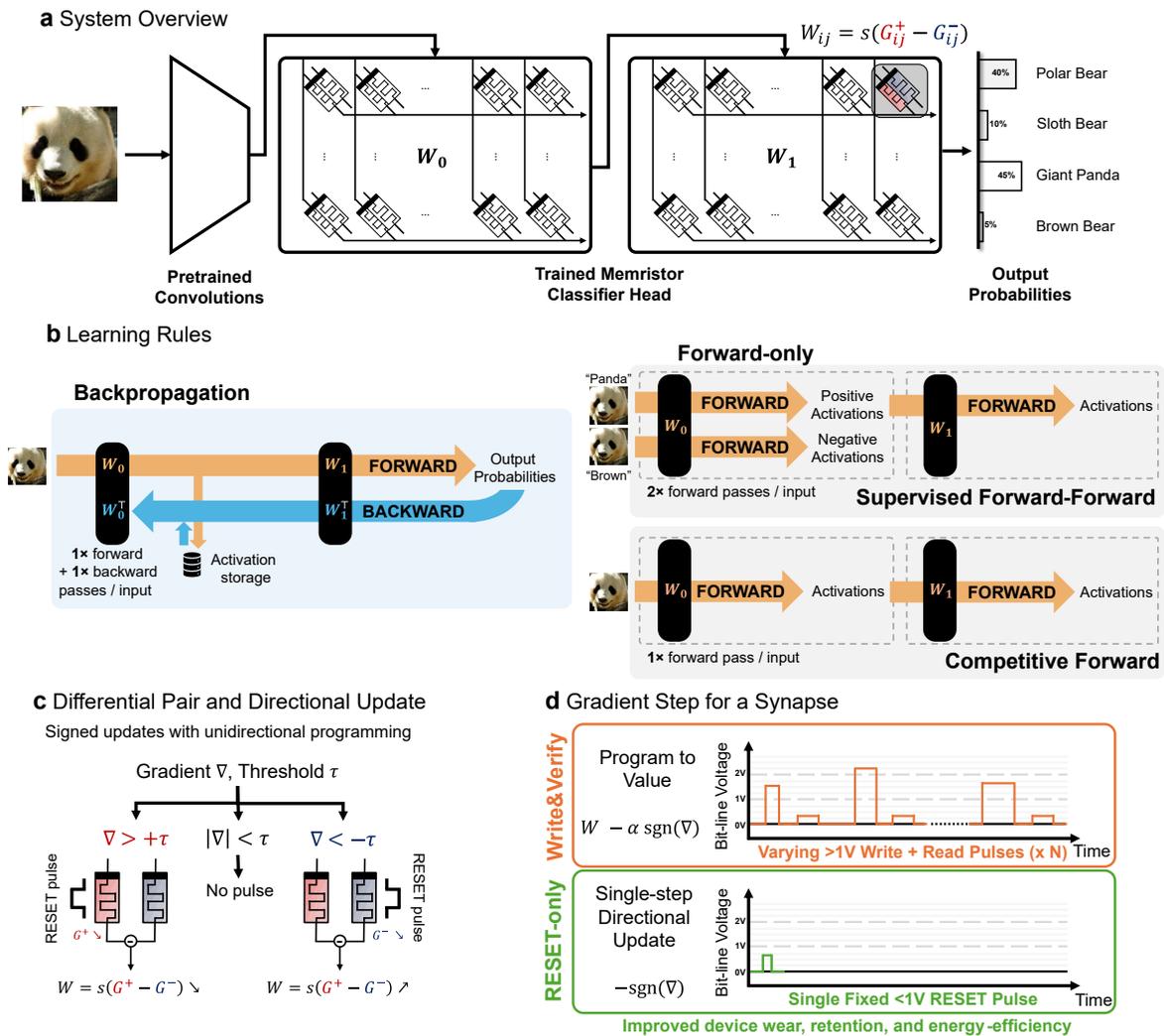}
    \caption{
    \textbf{Overview of sub-1 V reset-only learning and forward-only training in filamentary HfOx/Ti memristor arrays.}
\textbf{a.} Transfer-learning workflow used in this work: an ImageNet-pretrained convolutional backbone produces feature vectors that are classified by a memristor-implemented head \changed{trained through physical conductance updates} for four bear species. Synaptic weights are encoded by differential conductance pairs,
$W_{ij} = s \left( G_{ij}^{+} - G_{ij}^{-} \right).$
\textbf{b.} Learning-rule comparison. Backpropagation requires storing activations and propagating error signals backward (accessing transposed weights), whereas forward-only rules use inference-style forward passes with layer-local objectives. Supervised Forward-Forward uses positive and negative passes, while competitive forward uses a single forward pass with clustered neurons discriminating class labels.
\textbf{c.} Signed, selective weight updates with unidirectional device programming. A gradient estimate $\nabla$ is thresholded by $\tau$; when $|\nabla| > \tau$, a single sub-1~V RESET pulse is applied to one device of the differential pair to implement a sign-only update (RESET on $G^{+}$ decreases $W$; RESET on $G^{-}$ increases $W$).
\textbf{d.} Programming principle compared with conventional program-and-verify: rather than iteratively targeting a precise conductance state, the proposed regime targets only the update direction and applies one fixed-amplitude RESET pulse per triggered update, improving energy efficiency, endurance, and retention.
}
    \label{fig:principle}
\end{figure}

Here, we experimentally demonstrate a transfer learning task consisting of classifying ImageNet-resolution images of bears, using standard filamentary HfO$_x$/Ti devices (Fig.~\ref{fig:principle}a). We use forward-only training algorithms derived from Hinton’s Forward-Forward framework \cite{hinton2022forward}, which optimize layer-local objectives using only forward passes (i.e., inference-style MAC operations) (Fig.~\ref{fig:principle}b). We evaluate supervised Forward-Forward (SFF), which uses positive and negative examples and requires two forward passes per input, and a competitive, single-pass forward rule inspired by \cite{Papachrist24ClusterF} and named Competitive Forward (CF). The two  algorithms reach accuracies of 89.5\%  and 89.6\%, respectively on bear species classification, comparable to a backpropagation reference experiment using the same devices (90.0\%). 

In all experiments, we use a sub-1~V reset-only, single-pulse update primitive to update weights: 
each selected synapse receives one fixed-amplitude Reset pulse that induces a small, predominantly progressive conductance decrease. Signed learning is implemented with differential encoding,  $w\propto G^{+}-G^{-}$, by applying the Reset pulse to one device of the pair depending on the update sign (Figs.~\ref{fig:principle}c,d). For on-chip learning, reliable update direction and stability are more critical than precise conductance targeting, making such reset-only single-pulse updates a practical alternative to program-and-verify. Trained models retain accuracy for at least one month under ambient conditions, consistent with the strong retention of Reset-programmed states. Reset-only updates reduce programming energy by 460$\times$ relative to program-and-verify and require only 46\% more energy than inference-only operation.
%In this work, to accelerate multilayer experiments, we execute all array programming events in hardware while emulating MAC operations in software; we additionally demonstrate fully in-hardware training on a one-layer classifier.
\changed{
In this work, the single-layer perceptron experiment is fully in hardware: both analog MAC operations and sub-1~V conductance updates are performed on chip. To accelerate the larger multilayer experiments, we use a hardware-in-the-loop protocol in which all array reads and programming events are executed physically, while the MAC operations are emulated in software using conductances measured from the array.
}

%TC:ignore
%TC:endignore

\FloatBarrier

\changed{To the best of our knowledge, this Article presents the first large-scale (up to 8,064 devices) Forward-Forward-type experimental demonstration on memristors, the first report of month-scale accuracy stability after physical memristor-array training, and the first array-scale learning experiments based on sub-1~V reset-only physical conductance updates.}
Several experimental works have already demonstrated  in situ learning  on memristive hardware, using different strategies to approach the challenge of hardware backpropagation. Early demonstrations trained single-layer networks, which do not require backpropagation \cite{lin2016physical,alibart2013pattern,prezioso2015training,serb2016unsupervised,mao2022experimentally}. STELLAR avoided backpropagation by training only the second layer of a two-layer network \cite{zhang2023edge}. Other studies used a software backward pass \cite{li2018efficient, lin2024deep, liu2025error} or a Hopfield-type local activity-difference rule avoiding backpropagation but requiring iterative equilibria per input \cite{yi2023activity, park2025energy}. Ref.~\cite{van2024hardware} implemented a backward pass in hardware but on a minimal XOR task using six nanodevices. Out of these works, only ref.\cite{liu2025error} shows experimentally accuracy stability after on-chip learning (60 hours). When multi-week retention is reported (e.g., $\sim$48 days in \cite{zhang2023edge} or 30 days in \cite{esmanhotto2022experimental}), it pertains to off-chip-trained weights written via program-and-verify. These gaps motivate the sub-1~V reset-only regime adopted here, targeting both safe operating voltages and post-training stability within the same in situ learning protocol, and they further motivate Forward-Forward-type training as a route to multilayer learning without backward signals \cite{hinton2022forward}.  An alternative route is to learn in easier-to-write memories (SRAM \cite{wen2024fusion} or ferroelectric capacitors \cite{martemucci2025ferroelectric}) and periodically transfer weights to memristors, at the cost of considerable auxiliary memory.  A consolidated comparison of our work with prior experimental demonstrations of in situ learning in memristive hardware is provided in Suppl. Note~6.

\changed{
Preliminary, simulation-based versions of the Forward-Forward mapping and sub-1~V reset-only learning concept appeared in ref.~\cite{imbert2024forward}. The present Article presents the physical-array realization and system-level validation of the approach,  using up to 8,064 devices, on Forward-Forward and the more recent, single-pass competitive forward. These experiments allow us to report month-scale stability, device-population statistics, pulse-count analysis, energy accounting, and pulse-aware training adaptations.}
Together, the results of this Article chart a practical path from energy-efficient in-memory inference to on-chip learning using standard filamentary stacks operated safely below one volt.

\vspace{2cm}

\section*{Results}\label{sec:THR}

\subsection*{Sub-1~V memristor programming}

%TC:ignore
\begin{figure}[h]
    \centering
    \includesvg[width=0.8\linewidth]{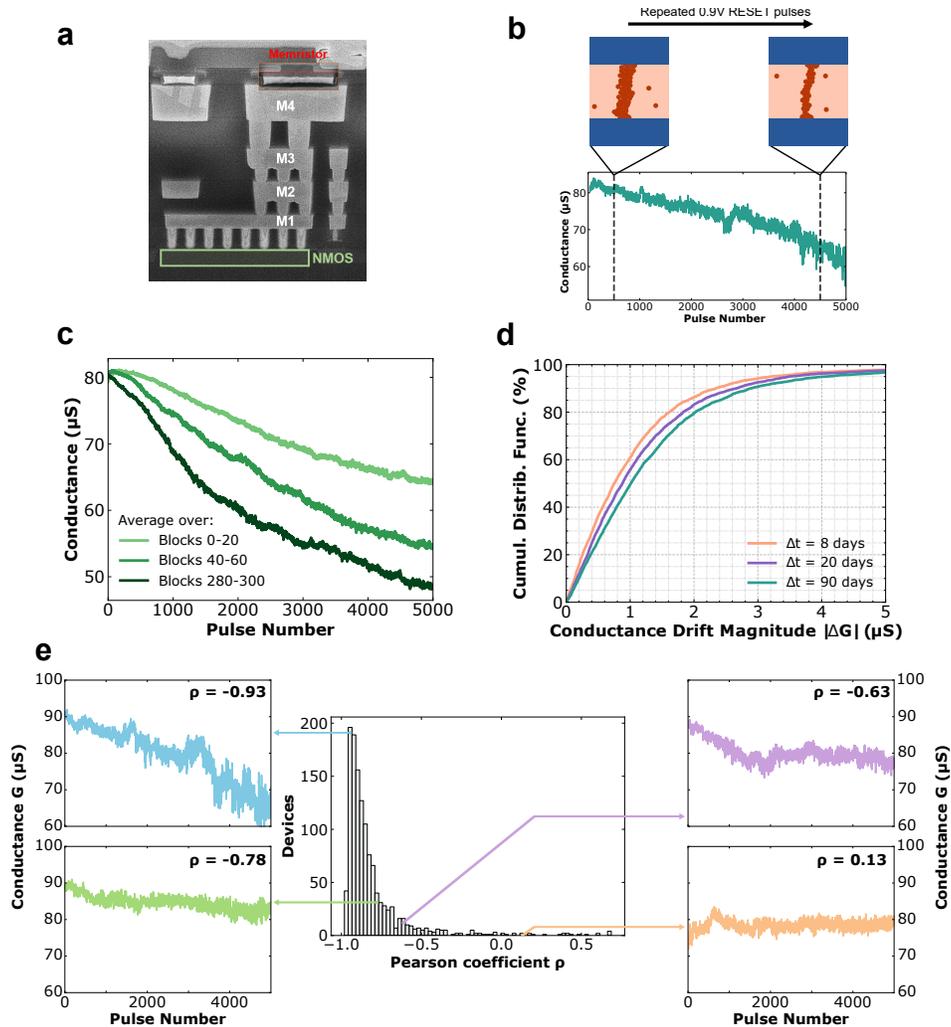}
    \caption{
    \textbf{Characterization of Sub-1~V reset regime of HfO$_x$/Ti memristors.}
    \textbf{a} Focused ion beam–scanning electron microscopy (FIB-SEM) image of a cut circuit showing memristor integrated over CMOS.
    \textbf{b} Proposed low-voltage reset procedure: No conductance is targeted, only its progressive decrease via the partial dissolution of an existing filament via repeated, same-low-voltage pulses.
    \textbf{c}~Endurance of the devices programmed with sub-1~V reset: the dynamics over blocks of  5,000 pulses stays similar after 300 blocks for a total of 1,500,000 pulses. The curves are averaged over four devices and over blocks 0-20, 40-60, and 280-300. 
    %The increase in memory window is faster from cycles 0 to 60 than from cycles 60 to 300.
    \textbf{d} Retention of 3,456 devices programmed with sub-1~V reset: 8 days after programming, 94.1\% drifted by less than 3~$\mu$S, and 90.7\% after 90 days.
    \textbf{e}  Distribution of the Pearson coefficient of conductance vs. pulse number curves for 1,268 memristors. Randomly-chosen examples of curves are presented for different Pearson coefficients.}
    \label{fig:weakreset}
\end{figure}
%TC:endignore

Memristor arrays excel at analog inference, but on-chip learning requires repeated, fine-grained weight updates, making device programming variability and wear rate central constraints. In filamentary devices, the common approach is to program conductance by controlling the compliance current during the set operation and using program-and-verify loops\cite{esmanhotto2022experimental,rao2023thousands}. This approach achieves precise targets but demands bit-line voltages exceeding 1~V and incurs multiple pulses per update, which is detrimental for device endurance. These conditions may be used in specially optimized few-shot learning contexts\cite{pallo2025chip} but are incompatible with frequent learning steps.

We instead exploit progressive filament dissolution using low-amplitude reset pulses. Short sub-1~V pulses applied to the bit line partially erode the conductive filament and monotonically decrease conductance by small increments (Fig.~\ref{fig:weakreset}b,c); device and characterization details are provided in Methods. Reset-only control is unidirectional. However, in nearly all memristor analog neural networks, each synaptic weight is encoded as a differential pair of conductances, \(w \propto G^{+}-G^{-}\) (this technique allows implementing signed weights). Therefore, in our case, a positive weight update is implemented by slightly resetting the ``negative'' device, and a negative update by resetting the ``positive'' device (Fig.~\ref{fig:principle}c). This scheme provides signed updates while keeping all programming \changed{below one volt}. 
Note that, throughout, ``sub-1~V'' denotes \changed{the magnitude of} the memristor terminal voltage (bit line to source line, \changed{$|V_\mathrm{BL}-V_\mathrm{SL}|$)}. In our 130-nm 1T1R arrays, we overdrive the word line (>1~V) to reduce access resistance; this does not raise \changed{$|V_\mathrm{BL}-V_\mathrm{SL}|$}, which stays <1~V during reset-only updates, and is unnecessary in advanced nodes (see Methods).

When training, to reduce energy and wear, and avoid the low-conductance regime where sub-1~V resets lose monotonicity, we restrict per-device updates with two rules. \emph{Selective updates:} apply an update only when the gradient magnitude exceeds a threshold~$\tau$. \emph{Sign-only pulses:} when triggered, ignore the gradient magnitude and apply exactly one sub-1~V reset to the appropriate device of the differential pair (a positive gradient pulses the ``positive'' device; a negative gradient pulses the ``negative'' device).

Operating in the sub-1~V reset-only regime brings two benefits. First, endurance is promoted because each update uses a single, low-voltage pulse. We stressed four devices with 1.5 million 0.9~V resets, inserting a full reset/set reinitialization every block of 5,000 pulses to return to low resistance. The progressive resistance increase is preserved (Fig.~\ref{fig:weakreset}c), which is the feature critical for learning. %Notably, the memory window widens with cycling, consistent with a reset ``wake-up'' effect.
\changed{Here, the ``memory window'' denotes the conductance span traversed during one fixed 5,000-pulse low-voltage reset trajectory, $G_{start}-G_{end}$. The widening seen in Fig.~\ref{fig:weakreset}c means that the same 0.9~V pulse train produces a larger conductance decrease after repeated reset/set conditioning cycles. We refer to this cycling-induced conditioning as a reset ``wake-up'' effect: the filament and its surrounding defect/trap configuration become more responsive to subsequent low-voltage reset pulses. The effect is strongest during the initial cycling blocks and then tends to saturate, as indicated by the smaller change between blocks 60 and 300.}

\changed{Second, retention is enhanced notably relative to states created by set with compliance-current programming. We interpret this improvement as a consequence of how the analog states are reached. In a conventional set/compliance or set-based Write\&Verify scheme, the target conductance is produced by a high-field, high-current set process. The resulting analog state can depend on a narrow, recently formed filamentary constriction and on a non-equilibrium distribution of oxygen vacancies and traps, so small relaxation processes or repeated reads can produce measurable conductance drift. In the sub-1~V reset-only regime, by contrast, devices are first placed in a low-resistance state and are then moved only in the reset direction. Each update slightly dissolves an existing filament rather than regrowing a new one. }
 %Second, retention is enhanced notably relative to states created by set with compliance current programming: starting from a thick, stable filament (low-resistance state) and only partially dissolving it yields states that drift less over time. 
  This high retention is seen in Fig.~\ref{fig:weakreset}d. We programmed 3,456 devices to various conductances ranging between 16 $\mu$S and 100 $\mu$S and measured the drift in their conductance after 8, 20, and 90 days. Fig.~\ref{fig:weakreset}d shows the cumulative probability distribution of the measured drift and confirms the high stability of the programmed states: 90.7\% of devices have drifted by less than 3 $\mu$S after 90 days. Suppl. Fig.~2 shows that the more conventional approach of programming by set under current compliance produces markedly higher drift within hours. The superior stability of reset-programmed analog states has also been observed using regular (>1~V) reset\cite{baroni2022low,pallo2025chip}.

These advantages come with characteristic non-idealities: Update magnitudes are stochastic from pulse to pulse due to the atomic-scale filament geometry and trap-assisted processes\cite{majumdar2021model}. The effective analog memory window  is also reduced compared with conventional program-and-verify: As the filament approaches full dissolution, a late-stage regime exhibits increased variability, which limits the practical number of reliable pulses between reinitializations. 

Device-to-device dispersion is also significant; nevertheless, the progressive, mostly monotonic behavior is observed consistently in all of the measured devices. To characterize the quality of the sub-1~V reset regime of a particular  device we compute the Pearson correlation coefficient between number of applied pulses and device conductance (see Methods). This coefficient measures how well the conductance vs. pulse number curve can be approximated by a straight line. Fig.~\ref{fig:weakreset}e shows the distribution of the Pearson coefficient of 1,268 devices, and for select values, a random example of conductance vs. number of pulses curves. Most devices have a Pearson coefficient close to -1, indicating a nearly linear relationship between conductance and pulse number. However, a non-negligible subset deviates from this trend, with some devices showing Pearson coefficients near zero or even positive. In the next section, we assess experimentally the impact of these imperfect devices on learning.

The sub-1~V reset regime can be optimized to a certain extent. We find that several aspects of the sub-1~V reset-only response can be tuned through forming conditions (Supplementary Note~1). In summary, a single fixed bias-forming pulse followed by set-reset ``wake-up'' cycles establishes a progressive, approximately monotonic reset slope. 
Still, the response remains stochastic to a large extent. In the remainder of this Article, we design and evaluate learning procedures that are explicitly matched to these programming characteristics and demonstrate experimentally that they are sufficient for stable, large-scale training on memristor arrays.

\FloatBarrier

\subsection*{Backpropagation-based learning with sub-1~V memristor programming}

%TC:ignore
\begin{figure}[h!]
    \centering
    \includesvg[width=0.8\linewidth]{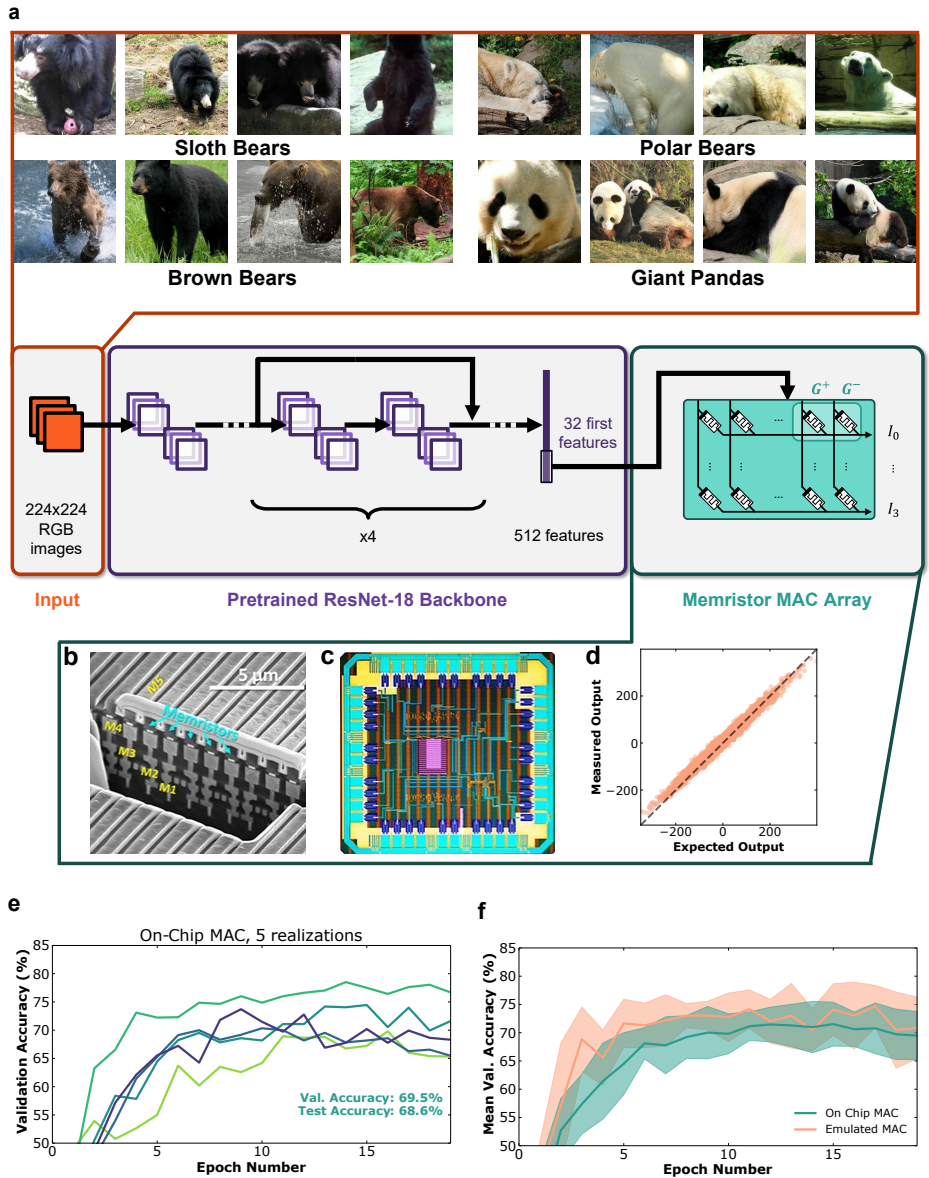}
    \caption{\textbf{Experimental transfer training of a memristor crossbar on bear classification using perceptron architecture. }
    \textbf{a} Topology of the trained architecture associating a ResNet pretrained on ImageNet with a memristor crossbar trained on bear species classification.
    \textbf{b} Focused ion beam–scanning electron microscopy (FIB-SEM) image of a memristor array integrated over CMOS.
    \textbf{c} Optical microscopy image of the multiply-and-accumulate (MAC)  memristor/CMOS integrated circuit.
    \textbf{d} Experimental validation of the MAC functionality with randomized weights: measured output as a function of result expected from software MAC on a variety of random inputs, after conversion of current to values (see Methods).
    \textbf{e} Evolution of validation accuracy over five experimental realizations of the training.  \changed{Both analog MAC operations and sub-1~V memristor programming updates are performed on chip.}
    \textbf{f} Comparison of the mean result of \textbf{e} with the mean results of five control experiments where only memristor updates are performed on-chip, and MACs are emulated by software to speed up experiments. 
        }
    \label{fig:backprop1}
\end{figure}
%TC:endignore

We target an edge-learning setting in which a cloud-pretrained backbone is adapted in situ via transfer learning rather than trained end-to-end. Concretely, we take an ImageNet-pretrained network, remove its final classifier, and replace it with memristor-implemented layers retrained on device to specialize the model (Fig.~\ref{fig:backprop1}a). In this proof-of-concept, the convolutional backbone is emulated in software. In a deployable system, it could be realized as an ultra-efficient fixed-weight front end that hardwires the convolutional synapses, trading programmability for extreme energy efficiency. Multiple demonstrations of such hardwired convolutional backbones have been demonstrated using register-transfer-level  synthesis \cite{miro2024772muj} or read-only memory \cite{chen2022yoloc,yin2023cramming,yu2025dsc}.

As a compact, ImageNet-resolution task, we classify four bear species (brown bear, sloth bear, polar bear, giant panda) using only a few thousand labeled images for fine-tuning (see Methods). The ImageNet-pretrained backbone yields a 512-dimensional feature vector at its penultimate layer. To match the input of our memristor arrays, reduce data movement, and keep the problem challenging for this proof-of-concept, we retain only the first 32 channels of this vector and train the memristor head on them. This 32-dimension input is used identically for train/validation/test across software baselines and hardware runs (see Methods; Supplementary Fig.~5 confirms that the 32-dimension task remains solvable but is harder than  using the full 512 dimensions).

Before studying forward-only learning approaches, we first perform reference experiments using conventional, backpropagation-based learning. We consider single-layer (perceptron) and two-layer memristor classifiers. Figs.~\ref{fig:backprop1}b,c show the hybrid CMOS/memristor ``MAC array'' circuit that we use to train the perceptron. It features a 32$\times$64 memristor array that performs multiply-and-accumulate (MAC) operations in analog using Ohm’s and Kirchhoff’s laws (see Methods). In differential encoding, columns for $G^{+}_{ij}$ and $G^{-}_{ij}$ are driven by $\pm x_jV_{read}$. The column current for class $i$ is therefore
\begin{equation}
I_i \;=\; \sum_j\!\left(G^{+}_{ij} - G^{-}_{ij}\right)x_jV_{read},
\end{equation}
and logits are read as $y_i=\kappa\,I_i$ with a fixed gain $\kappa$. Fig.~\ref{fig:backprop1}d  shows the measured output of the MAC array versus the expected software MAC, confirming high accuracy over a wide input/output range.

To perform a learning experiment, for each  batch of data, 
%${\mathbf{x}, \mathbf{\hat{y}}} \in \mathcal{B}$, 
we perform the forward pass experimentally, compute the gradients in software based on the results of the forward pass (see Methods) %$\mathbf{y} = f(\mathbf{x})$ using a cross-entropy loss
%\begin{equation}\label{eq:grad_bp}     g_{i j} = \frac1{|{\mathcal{B}}|} \sum_{\mathbf{x}, \mathbf{\hat{y}} \in \mathcal{B}}(y_i - \hat{y}_i) \cdot x_j \end{equation}
and then adjust the conductances based on the principles described in the previous paragraph.  Experimentally (five independent runs), the perceptron reaches a mean final accuracy of $69.5\%$ (Fig.~\ref{fig:backprop1}e). The experiments have high variation as the accuracy ranged between 65.3\% and 76.7\% (the standard deviation is 4.3\%).

%TC:ignore
\begin{figure}[h]
    \centering
    \resizebox{\columnwidth}{!}{%
    \includesvg[width=\linewidth]{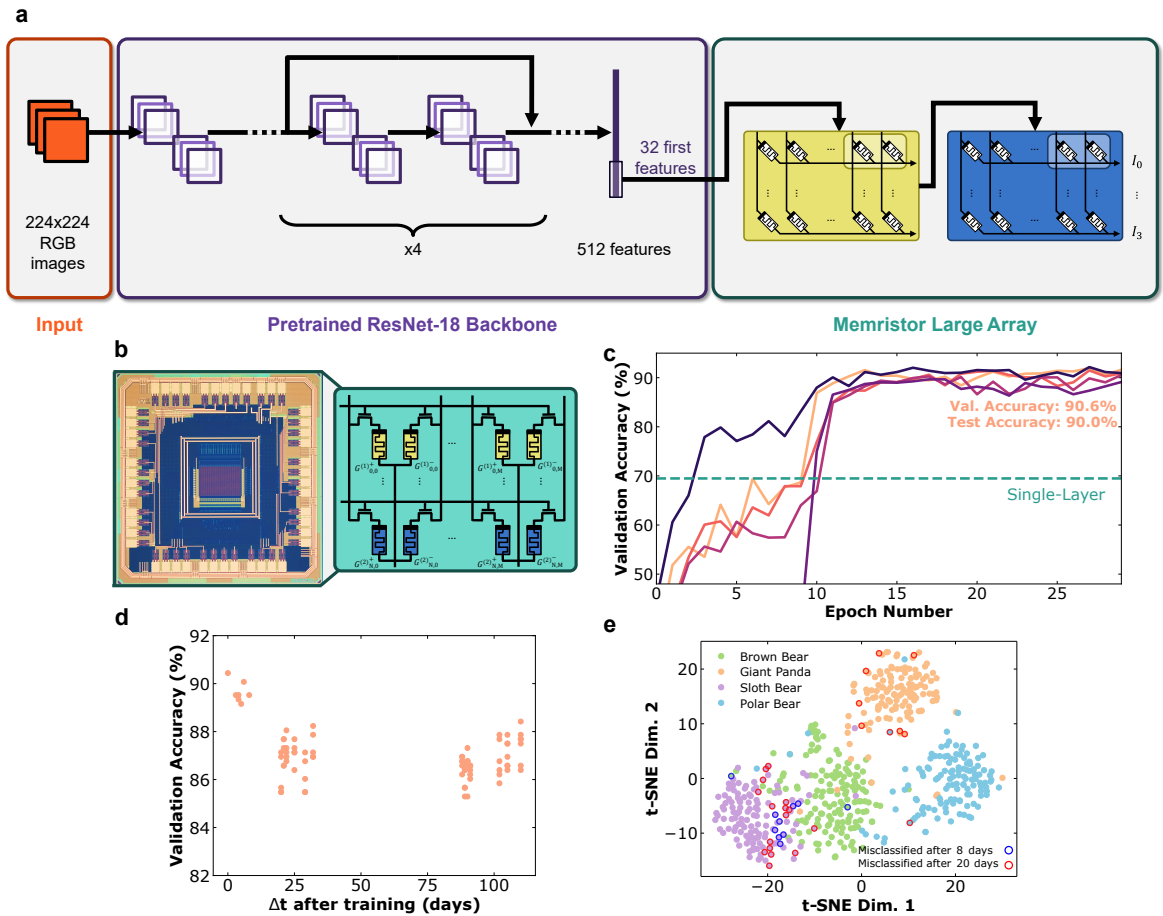}
    }
    \caption{\textbf{Experimental transfer training of a memristor crossbar on bear classification using a multilayer perceptron architecture and backpropagation. }
    \textbf{a} Topology of the trained architecture associating a ResNet pretrained on ImageNet with two memristor crossbars.
    \textbf{b}  Optical microscopy image and simplified schematic of the ``large array''  memristor/CMOS integrated circuit.
    \textbf{c} Evolution of validation accuracy over five experimental realizations of the training. 
     \changed{All conductance reads and programming updates are performed on the physical Large Array, while MACs are emulated in software using the measured conductances.}
    %All memristor updates are performed on-chip, MACs are emulated by software to speed up experiments (unlike Fig.~\ref{fig:backprop1} where they are performed on chip).
    \textbf{d} Evolution of the validation accuracy over time once training is finished for one circuit.
    \textbf{e} t-SNE representation of the bear test dataset. The samples initially classified correctly and then incorrectly after 8 and 20 days are marked.
       }
    \label{fig:backprop2}
\end{figure}
%TC:endignore

\FloatBarrier

For the two-layer network (Fig.~\ref{fig:backprop2}a), we use the ``Large Array,'' a 128$\times$64 1T1R hybrid CMOS/memristor chip with direct analog device access but no on-chip MAC (Fig.~\ref{fig:backprop2}b; Methods). To accelerate multilayer experiments, we adopt a hardware-in-the-loop protocol: MACs are emulated in software while all weight reads/writes occur on hardware. At each training step we (i) read all devices and map them to weights using $W=s(G^{+}-G^{-})$; (ii) run batched inference and compute gradients in PyTorch; (iii) apply at most one sub-1~V reset to the appropriate device of each selected differential pair (sign-only, thresholded); (iv) re-read to refresh $W$. With this technique all device-level effects of the sub-1~V reset regime (stochastic step sizes, limited analog window, drift) are still done in hardware, while greatly reducing runtime. We verified on the perceptron that on-chip MAC and hardware-in-the-loop experiments yield matching learning curves and final accuracy (Fig.~\ref{fig:backprop1}f).

\changed{We found that simultaneous training of all layers, as usually done in software, is poorly matched to stochastic reset-only updates (Suppl. Fig.~6), because update stochasticity is then propagated across layers. A random conductance change in an early layer perturbs the activations used to update the next layer, while stochastic changes in the later layer alter the objective seen by the earlier layer. Because all layers are changing at once, these perturbations compound through the network and require more programming pulses. 
We therefore adopted a layer-wise schedule in which only one layer is plastic at a time; the other layer is frozen, making the local input distribution and objective more stationary and improving convergence under hardware nonidealities.}
%We found when training multilayer classifiers that training all layers simultaneously, as is usually done in software, amplifies device noise and stochastic update variability (see Suppl. Fig.~5). We therefore adopted a layer-wise schedule: first train the layer closest to the output, then freeze it and train the preceding layer.  This simple schedule consistently improved convergence under hardware non-idealities.

Experimentally, the two-layer network (multilayer perceptron, MLP) attains 90.0\%,  with markedly tighter standard deviation (1.1\% over five runs)   than in the perceptron case (Fig.~\ref{fig:backprop2}c). For comparison, software baselines of perceptron and two layer neural network (same architecture, floating-point training) achieve $93.3\%$ and $93.8\%$, respectively. The larger hardware-software gap for the perceptron is noteworthy: Despite its simplicity, the single layer is more sensitive to stochastic, unidirectional updates and device-to-device dispersion, whereas the hidden layer in the MLP provides representational redundancy that absorbs programming noise and limited analog dynamic range. This behavior underscores that learning dynamics under sub-1~V reset constraints differ from conventional floating-point training, as in software implementations, perceptrons typically show easier and more robust convergence than MLPs. The three percentage point gap in the MLP case is significant, but it is still appreciably low when considering the highly stochastic nature of the sub-1~V reset regime.

As expected from the characteristics of the sub-1~V reset, the trained neural network shows remarkably stable accuracy, with no significant degradation one week after training (Fig.~\ref{fig:backprop2}d) and three percentage points after one and three months. This result mirrors the remarkable stability of sub-1~V-reset-programmed memristor states observed in Fig.~\ref{fig:weakreset}d. 

Fig.~\ref{fig:backprop2}e investigates which are the few bear samples that \changed{became}  misclassified after 8 or 20 days although they were correctly classified after training. For this purpose, we plotted a t-distributed stochastic neighbor embedding (t-SNE) representation of our test dataset (see Methods), and we marked the samples that became misclassified in time. We observe that these few samples typically lie on the edges of the t-SNE blobs representing the different bear species. This observation suggests that they are \changed{borderline}, harder-to-classify samples. A similar observation has been made in the literature when looking at errors on a poorly illuminated, solar-powered near-memory-computing circuit \cite{jebali2024powering}.

\FloatBarrier
\subsection*{Forward-pass-only learning with sub-1~V memristor programming}

%TC:ignore
\begin{figure}[h! ]
    \centering
    \includesvg[width=\linewidth]{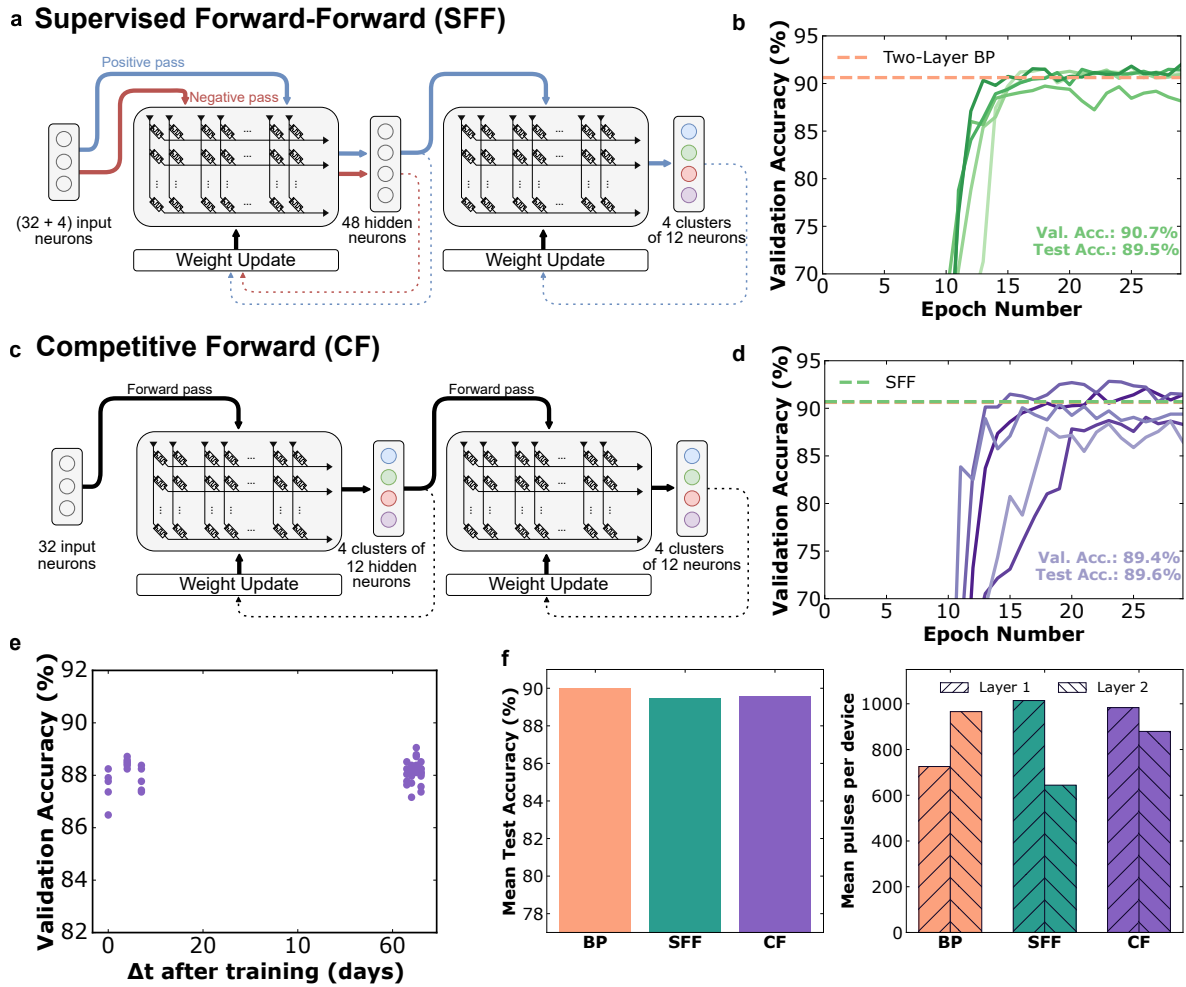}
    \caption{\textbf{Experimental transfer training of a memristor crossbar on bear classification using forward-only training approaches. }
    \textbf{a} Illustration of our Forward-Forward approach using ``positive'' and ``negative'' examples and a second memristor layer with competitive clusters.
    \textbf{b} Evolution of validation accuracy on bear classification over five experimental realizations of Forward-Forward training. 
   \changed{The hardware-in-the-loop methodology is the same as in Fig.~\ref{fig:backprop2}: physical array reads/writes and software-emulated MACs.} 
    %Methodology for experiments is the same as Fig.~\ref{fig:backprop2}.
    \textbf{c} Illustration of our competitive forward approach using two memristor layers with competitive clusters.
    \textbf{d} Evolution of validation accuracy on bear classification over five experimental realizations of competitive forward training.
    \textbf{e} Evolution of the validation accuracy over time once training is finished for one circuit trained with competitive forward. 
    \textbf{f} Comparison of mean test accuracy, and mean number of sub-1~V reset pulses per device in our experiments employing backpropagation, Forward-Forward, and competitive forward training.}
    \label{fig:forward}
\end{figure}
%TC:endignore

In the multilayer backpropagation experiments reported in the previous section, all array write operations are performed on hardware, while the surrounding control is emulated. Building a fully integrated system around backpropagation, however, is challenging: it requires a backward pass (transpose-array access or weight reuse with reversing signal flow), storage of layer activations, and additional on-chip compute, which together impose substantial circuit and energy overhead. These challenges motivate forward-pass-only alternatives that avoid any backward signals. 

We therefore evaluate two hardware-friendly schemes. The first is Hinton’s Forward-Forward  algorithm \cite{hinton2022forward}. Each layer computes a scalar ``goodness'' of its activations,   which is the sum of the squared activations of its neurons (see Methods). Training uses only forward passes on positive (data paired with the true label token) and negative (data paired with an incorrect label token) examples to maximize goodness on positives and minimize it on negatives. This is easily achieved using local weight updates requiring no backpropagated errors (see Methods, eq.~\eqref{eq:sfffinal}).

For readout, Hinton recommends using a large softmax head to extract output, spanning over several neural network layers, but this approach is poorly matched to sub-1~V reset. As shown earlier, shallow readouts degrade under hardware non-idealities. Second, it would require storing the activations of several layers, making it complicated to operate learning in a pipelined fashion (i.e., present a new input at the first layer, while the previous input is processed on the second layer). 

To replace this softmax head, we use a memristor array acting as a bank of clusters (prototypes), one per class, each with 12 neurons (Fig.~\ref{fig:forward}a). The idea is that during training, we aim at maximizing the goodness of the target cluster and minimizing the goodness of the other clusters. Conveniently, we can use the same goodness definition  as the first layer, and the same local rule (see Methods, eq.~\eqref{eq:cffinal}). 

We implement Forward-Forward on memristor arrays using the same sub-1~V, sign-only, selective update rule as above and train layers sequentially. Unlike in the backpropagation case, we start training from the input side: As information is transmitted only in the forward direction in the Forward-Forward algorithm, no other sequence is possible. Experimentally (five runs), supervised Forward-Forward  achieves a mean test accuracy of 89.5\% (the standard deviation over five runs is 1.4\%), within experimental uncertainty of the two-layer backpropagation baseline (90.0\%  with 1.1\% standard deviation,   Fig.~\ref{fig:forward}b). In floating-point software, the two methods likewise match ($\sim\!93.8\%$). Consistent with this parity, the mean pulse count per device is similar across supervised Forward-Forward and backpropagation. However, the layer-wise distribution differs: in backpropagation the second (output) layer accumulates the most pulses, whereas in supervised Forward-Forward the first (input) layer does so (Fig.~\ref{fig:forward}f). This inversion follows directly from the training schedules: backpropagation trains layers from output to input, supervised Forward-Forward from input to output,  so in both cases the layer trained first receives the largest share of updates, while the second-trained layer primarily fine-tunes the network.

An important challenge with supervised Forward-Forward is the need to build positive and negative training examples, and the cost to perform two corresponding forward passes.  To mitigate this issue, we built a network by stacking two cluster-based layers, similar to the one we use for readout in our supervised Forward-Forward experiments: we obtain 
what we call competitive forward (CF) learning, a purely forward, single-pass scheme built from stacked, all-to-all cluster layers  (Fig.~\ref{fig:forward}c). Each layer learns a set of competing prototypes; with depth, prototypes become more selective, and the final layer’s clusters correspond to classes. Updates are local, use the same goodness equation as in supervised Forward-Forward, and the same selective, sign-only, single-pulse rule as in our other experiments, and layers are trained sequentially from input to output. The name ``competitive forward'' comes from the work of ref.~\cite{Papachrist24ClusterF}, which proposes a similar approach in software. However,  whereas  ref.~\cite{Papachrist24ClusterF} instantiates competitive clusters on the convolutional feature maps solely, we instead implement them as fully connected classifier heads, avoiding large softmax heads and aligning naturally with 1T1R crossbar topology and sub-1~V reset constraints.

As with supervised Forward-Forward, we train competitive forward layer by layer from input to output (see Methods). Note that we inverted the sign of the goodness of the first layer (see Methods): this way, the first layer is driven to decrease the activity of the target cluster and increase the activity of the complementary clusters; the output layer  then concentrates activity on the correct class. 
This sign asymmetry was not beneficial in floating-point software, but on hardware it reduces the pulse burden on the first layer (the one trained first), helping to avoid late-stage variability under sub-1~V reset and yielding more stable convergence \changed{in the case of the animals dataset}.

Fig.~\ref{fig:forward}d shows that across five experiments, competitive forward learning  matches supervised Forward-Forward performance (mean test accuracy of 89.6\%, with a standard deviation over five runs of 1.8\%). A statistical test (see Methods) shows the \changed{differences} in test accuracy observed between our experimental realization of competitive forward, supervised Forward-Forward, and the backpropagation experiments are not statistically significant, suggesting that the forward-only approaches are an excellent hardware-friendly alternative to backpropagation for the transfer learning task investigated here.

The mean number of pulses per device is similar between competitive forward and supervised Forward-Forward (Fig.~\ref{fig:forward}f), with values around or below 1,000. Considering that we have demonstrated endurance of at least 1.5M pulses (Fig.~\ref{fig:weakreset}c), this result means that the learning process can be repeated more than 1,500 times using the same memristor array.

Finally, Fig.~\ref{fig:forward}e shows that the accuracy of a network trained by competitive forward  is highly stable over two months. Unlike in the backpropagation case (Fig.~\ref{fig:backprop2}d), no accuracy drop is seen (note that this experiment is performed with a network with an initial accuracy of 88\%, i.e. lower than average). We attribute this even higher stability to the larger size of the forward-only network, and in particular, the final layer naturally features more redundancy than in the backpropagation case due to its cluster nature.

\begin{figure}[t]
    \centering
    \includesvg[width=0.75\linewidth]{figures/6a_read_disturb.svg}
    \\[0.5cm]
    \includesvg[width=0.75\linewidth]{figures/6b_alt_trainings.svg}
    \caption{
    \changed{\textbf{Read-disturb stability and reprogrammability of the trained memristor network.}
    \textbf{a} Trained-model read-disturb experiment and matched Write\&Verify comparison. After programming, all memristors were read every 2~s for 25~h, and test accuracy was recomputed every 200~s. The reset-only curve corresponds to a competitive forward network trained directly with sub-1~V reset-only updates. The Write\&Verify baseline uses the final weights of the same competitive forward run, quantized to six conductance levels and programmed with a three-trial set-based Write\&Verify procedure. Curves show a moving average over 50 reported accuracy values; shaded regions indicate the corresponding standard deviation.
        \textbf{b} Repeated hardware reprogramming of the same Large Array for two different four-class transfer-learning tasks. The network was trained with competitive forward learning, alternating between the bear and bird tasks. Both tasks used identical hyperparameters. Between tasks, all memristors were globally reinitialized by hard RESET followed by SET; no device selection, remapping, or task-specific hyperparameter optimization was performed.}
    }
    \label{fig:practical_stability}
\end{figure}

\changed{
The stability measurements above were performed over long times, but with only sparse read operations between accuracy evaluations. In an edge inference system, the trained weights would be read repeatedly. We therefore performed an aggressive read-disturb experiment on a competitive forward network trained with sub-1~V reset-only updates. After training, all programmed memristors were read every 2~s for 25~h using the same read conditions as in the learning experiments, corresponding to approximately $4.5\times10^4$ read operations per memristor. The test accuracy was recomputed every 200~s. Under this protocol, the reset-only trained model shows no systematic degradation: the moving-averaged test accuracy remains within roughly half a percentage point of its initial value throughout the experiment (Fig.~\ref{fig:practical_stability}a).
}

\changed{
To compare against a conventional programming baseline at the trained-model level, we used the final weights of the same competitive forward training run, quantized them to six conductance levels, and programmed them with a three-trial set-based Write\&Verify procedure. The quantized software model reached 89.71\% test accuracy, compared with 90.62\% before quantization, and the physically programmed Write\&Verify model reached 90.07\% immediately after programming. However, under the same repeated-read protocol, the Write\&Verify baseline progressively lost nearly 3 percentage points of test accuracy after 25~h (Fig.~\ref{fig:practical_stability}a).
}

\changed{
We also tested whether the same array can be reprogrammed for different inference tasks. Using the same competitive forward architecture and identical hyperparameters, we alternated hardware training between the original four-class bear task and a four-class bird task. Each task was trained for 30 epochs, with the last layer trained only during the final 10 epochs. Between tasks, all memristors were globally reinitialized by hard RESET followed by SET; no device selection, remapping, or task-specific hyperparameter optimization was performed. The same physical array repeatedly recovered high validation accuracy on both tasks, with no monotonic degradation across task switches (Fig.~\ref{fig:practical_stability}b). This experiment demonstrates reprogrammability of the memristor head for different inference tasks. It should not be interpreted as information-preserving continual learning, because each task switch intentionally erases the previous model.
}

In the competitive-forward experiments on the large array, each programming pulse consumed 35~pJ on average. 
%With the MAC-array, which uses devices optimized for lower-voltage operation, the per-pulse energy drops to 0.84~pJ; 
\changed{To estimate the energy of the same pulse sequence on our lower-voltage platform, we used the parameters of the MAC array, i.e., the 32$\times$64 2T1R circuit used for the fully in-hardware perceptron experiment (Figs.~\ref{fig:backprop1}b-d; Methods). This platform was fabricated with a newer HfO$_x$ integration flow optimized for lower-voltage programming and uses 0.62~V, 30~ns reset pulses (Suppl. Table~1). With these MAC-array pulse parameters, the per-pulse energy is 0.84~pJ.}
By contrast, a program-and-verify baseline requires 387~pJ per update (see Methods). Thus, optimized devices reduce energy by 42 relative to the large array conditions and by 460 relative to program-and-verify. Aggregated over the full competitive-forward run, resets consumed 251~$\mu$J in our experiments; the same pulse counts on the optimized devices would total 6.0~$\mu$J (see Methods).

Our lab-bench reads used 15~$\mu$s integrations, which would lead to 21~mJ of read energy for the competitive-forward run. In a fully integrated implementation in an advanced CMOS node with faster sensing, these overheads should be far lower. Using the efficiency reported in ref.~\cite{xue2021cmos}, executing all MACs (including ADC) for the run would consume  13~$\mu$J (see Methods), which is twice as high as the  energy of sub-1~V reset pulses with the optimized devices:  because sub-1~V resets are so inexpensive, implementing competitive-forward in hardware  only adds modest overhead relative to inference-only designs. These extremely low-energy budgets are compatible with stringent edge power budgets. 
\changed{This comparison is array-level; Supplementary Note~5 further includes a projected local-controller cost, which can exceed the optimized reset energy but remains far below the program-and-verify programming cost.}

\FloatBarrier

\section*{Discussion}

\changed{
Our experiments establish that sub-1~V reset programming can support physical memristor-array learning of standard filamentary memristors. 
}
%Our experiments establish that on-chip learning is viable with sub-1~V reset programming of standard filamentary memristors. 
Competitive forward learning stands out as the most hardware-friendly approach, in terms of implementation and it leads to accuracy on par with backpropagation on our bear classification example. It also shows month-long accuracy stability, even higher than the already high accuracy of the backpropagation trained neural network,  as competitive forward learning uses a second layer with more redundancy than backpropagation.

%The experimental demonstrations presented in this Article target an ImageNet-resolution, four-class bear task. 
\changed{
The experimental demonstrations presented in this Article use an ImageNet-resolution, four-class bear task as a controlled physical-array benchmark. This setting provides a nontrivial transfer-learning problem while keeping the feature representation, array size, pulse budget, and measured-update constraints identical for layer-wise backpropagation, supervised Forward-Forward, and Competitive Forward. It therefore tests the central hardware-learning question of this work: whether standard filamentary HfO$_x$/Ti memristor arrays can support stable, low-voltage, pulse-aware learning with forward-only rules.
To connect this physical-array benchmark to a sensor-edge deployment scenario, Supplementary Note~5 analyzes a complementary electrocardiogram  (ECG) arrhythmia-classification pipeline in which the input is a low-bandwidth sensor signal, the fixed front end is compact and expressible as memristor MAC operations, and the adaptive classifier uses the same reset-only learning primitive. In that case, the front-end and classifier costs can be included in a common MAC-based energy accounting.
}

To assess generality and enable ablations that would be impractical on hardware, we also used device-calibrated simulations that replay measured sub-1~V reset trajectories from 1,268 devices: each synaptic weight is mapped to a differential pair, and a sign-only update advances exactly one pulse along the recorded trace, reproducing stochastic step sizes without parametric fitting. On the bear task, these simulations closely map the hardware learning curves and final accuracies (Supplementary Note~3\changed{, section ``Device-calibrated simulations of the experiments''}), supporting their use as a faithful proxy. In most cases experiments slightly outperform the device-calibrated simulations (Suppl. Table~3), which we attribute to the simulator’s finite pool of measured trajectories. Some traces must be reused across multiple simulated synapses, yielding a pessimistic estimate. Using the same simulator on the standard handwritten digit classification dataset MNIST (Supplementary Note~4), we observe qualitatively similar behavior and a comparable, modest gap to floating-point software: layer-wise backpropagation, supervised Forward-Forward, and competitive forward all perform well, with competitive forward consistently edging supervised Forward-Forward while retaining a single forward pass. \changed{
By contrast, simulations on CIFAR-10, a harder ten-class task, show a clearer advantage of supervised Forward-Forward over competitive forward (Suppl. Note~3, section ``Scaling with feature dimensionality and task complexity'').
}

A central takeaway is that the learning dynamics under low-voltage reset differ materially from conventional software training. Unidirectional, stochastic, pulse-quantized updates and a limited analog window shift the optimization objective from “precise gradient following” toward “minimizing effective pulse usage while preserving useful signal”.
In practice, this leads to design choices that are uncommon in software: selective (thresholded) updates, layer-wise training schedules, and classifier heads that offer representational redundancy (e.g., cluster layers) to absorb device variability.
Neuroscience simulations corroborate this last result, showing that redundancy allows reaching \changed{performance} on par with reconfigurability of synaptic connections\cite{hiratani2018redundancy}, which is non-trivial in silicon hardware. Furthermore, because of the sub-1~V reset-only scheme, we use one-bit (sign-only) updates.
Although motivated by hardware considerations, our approach is reminiscent of the ``signSGD'' optimizer introduced in software contexts \cite{bernstein2018signsgd}. Interestingly, to some extent, this optimizer appears as a special case of the Adam optimizer. Typically, Adam relies on first $m_t$ and second $v_t$ moments of the gradient $\nabla_t$, for an update $\propto m_t / \sqrt{v_t}$. However, without momentum we have $m_t = \nabla_t$ and $v_t = \nabla_t^2$, therefore $m_t / \sqrt{v_t} = \nabla_t / \sqrt{\nabla_t^2} = \mathrm{sgn}(\nabla_t)$.
This optimizer can outperform the regular Adam in some scenarios\cite{bernstein2024old,balles2020geometry}.

These choices reduce energy consumption and device wear and keep devices out of the late, high-variability dissolution regime. They also motivate pulse-aware training rules: regularize update counts, cap per-weight pulse budgets, and control update probability rather than amplitude. This algorithm–device co-design departs from standard learning-rate/weight-decay tuning and matches the quantized, unidirectional, stochastic nature of sub-1~V resets.

\changed{
This pulse-aware view is quantified by the accuracy-pulse Pareto analysis in Supplementary Note~3 (section ``Accuracy-pulse Pareto analysis''). In that analysis, hyperparameters are optimized jointly for validation accuracy and for the average number of sub-1~V reset pulses per active device. The high-accuracy operating points used in the hardware experiments lie in the sub-1000-pulse-per-device regime, below the \(\sim\!2{,}000\)-\(3{,}000\)-pulse range where low-voltage reset trajectories become markedly more irregular. Thus, the objective is not to maximize the number of updates, but to reach useful accuracy before the effective analog reset window is consumed.}

\changed{Several reasons contribute to the relative resilience of the experiments to  device mismatch. First, the relevant device property is not identical conductance change per pulse, but a statistically reliable update direction. Most devices show a progressive, approximately monotonic reset trajectory  (Fig.~\ref{fig:weakreset}e). Second, the update rule is thresholded and sign-only: small gradients are suppressed, and selected synapses receive one fixed pulse rather than an amplitude proportional to the gradient. Finally, multilayer hidden representations and multi-neuron class clusters distribute evidence over many synapses, reducing the impact of individual weak devices. 
This robustness is not unlimited: the device-imperfection sweep in Supplementary Note~4 shows that the most irregular reset trajectories cause accuracy losses in forward-only learning. In this severe-imperfection regime, layer-wise backpropagation can outperform the forward-only algorithms. The likely reason is pulse efficiency: backpropagation reaches useful solutions with fewer effective update cycles, while forward-only local objectives can require more pulses, which is costly for devices with a short monotonic reset range. This result confirms that limiting the number of programming pulses is key to high-quality in situ learning.
%This robustness is not unlimited: The device-imperfection sweep in Supplementary Note~4 shows that the most irregular reset trajectories cause accuracy losses in forward-only learning. When using devices in such severe-imperfection regime, layer-wise backpropagation can outperform the forward-only algorithms. The likely reason is pulse efficiency: backpropagation reaches useful solutions with fewer effective update cycles, while forward-only local objectives can require more pulses, which is costly for devices with a short monotonic reset range. This result confirms that limiting the number of programming pulses is a key for quality in-situ learning.
}

\changed{
A limitation of reset-only programming is that it is not, by itself, an unlimited continual-learning mechanism. Differential encoding allows signed weight updates because either $G^{+}$ or $G^{-}$ can be reset; however, every programming event decreases one conductance, so the common-mode conductance of a pair decreases with cumulative updates. If learning continues indefinitely, devices eventually approach the low-conductance regime where the effective analog window is exhausted and sub-1~V reset becomes less monotonic. %A work-around would be a refresh operation, where both devices and the device with the appropriate sign is reprogrammed to recreate  \(\Delta G=G^{+}-G^{-}\).
A possible workaround is information-preserving recentering: read the current differential value \(\Delta G=G^{+}-G^{-}\), reinitialize both devices to a high-conductance common-mode state, and then reset the device required to recreate the sign and magnitude of \(\Delta G\). %Such a refresh would add occasional higher-voltage or verify-assisted operations and was not implemented here.
} 
\changed{ The repeated bear-bird reprogramming experiment in Fig.~\ref{fig:practical_stability}b can also be interpreted within this framework. It demonstrates that the same physical array can be erased, reinitialized, and retrained for different inference tasks without observable degradation across the tested sequence. 
}

Our experiments use small minibatches (size 16) to stabilize updates. Because SFF and competitive-forward rules are local and we train layers sequentially, minibatching is much simpler than in backpropagation: to compute a layer’s weight updates, we only need that layer’s pre- and post-synaptic activations ($x_\ell$ and $h_\ell$). 
In practice, we buffer $x_\ell$ and $h_\ell$ for the 16 samples in a small working memory, accumulate sufficient statistics over the batch, and then apply the sign-only updates (see Methods, eqs.~\eqref{eq:sfffinal} and~\eqref{eq:cffinal}). 
The required working memory is moderate \changed{because it scales with the number of neurons in the active layer rather than with the number of synapses, unlike approaches that require synapse-sized auxiliary memory}
%The required memory is moderate as it scales with the number of neurons per layer, rather than the number of synapses as other works using auxiliary memory
\cite{wen2024fusion,martemucci2025ferroelectric}.  Supplementary Note~3 \changed{(section ``Simulation study of the effect of batch size'')} analyzes batch-size trade-offs, indicating that micro-batches (size smaller than 16) recover most of the stability benefits while further reducing storage. 

At the device level, the sub-1~V reset-only regime already delivers multiple advantages: safe sub-1~V operation, fewer pulses per update, improved endurance, and superior retention relative to set+compliance states. The remaining stochasticity, however, explains the few-percentage-point gap between multilayer hardware and software baselines, and the larger gap for single-layer perceptrons. Materials and stack engineering aimed at stabilizing filament geometry and trap landscapes, as well as pulse-shape optimization and modest circuit-level redundancy (e.g., small ensembles per weight or read averaging), are promising paths to close this gap without sacrificing the benefits of low-voltage operation.
\changed{
The ECG energy accounting in Supplementary Note~5 illustrates the resulting system-level shift: with conventional program-and-verify, learning would remain dominated by memristor programming, whereas sub-1~V reset-only updates reduce programming to a small term and move the main optimization target toward integrated MAC/read energy and the local learning controller.
}

Overall, our results chart a concrete route to practical, energy-efficient on-chip learning: adopt low-voltage reset programming; embrace pulse-aware, forward-only training rules that minimize updates; and provision minimal, local working memory where needed. With continued co-optimization across devices, circuits, and algorithms, including hybrid near-array memories and pulse-budgeted learning rules, the accuracy-energy envelope of sub-1~V memristor learning should improve further, enabling adaptive edge intelligence beyond the inference-only regime.

%TC:ignore

\FloatBarrier

\section*{Acknowledgements}

This work benefited from France 2030 government grants managed by the French National Research Agency (ANR-22-PEEL-0010, ANR-22-PEEL-0013, ANR-23-PEIA-0002). 
This project was provided with computing HPC and storage resources by GENCI at IDRIS thanks to the grant 2025-AD011016471 on the supercomputer Jean Zay's A100-H100 partition.
The authors would like to thank Olivier Faure, Julie Grollier, Louis Hutin, Jonathan Miquel, and David Novo for discussions and invaluable feedback.  A large language model (OpenAI ChatGPT) was used for copyediting parts of this manuscript.

\section*{Author contributions statement}
A.R. developed the code base of the project and performed the learning experiments, with assistance from M.A.I. 
M.H.D. proposed and developed the hardware adaptations of the Forward-Forward algorithms with J.D.A.-M. and F.M.
T.D. proposed cluster-based forward-only learning and performed all bear-dataset and MNIST simulations.
B.I. performed the preparatory simulations for hardware Forward-Forward algorithms.
A.R. and T.H. developed the initial backpropagation study.
D.B., A.R., and T.D. curated the bears dataset.
K.E.H. designed the hybrid CMOS/memristor test chip.
E.V. led the fabrication of the hybrid CMOS/memristor test chip.
C.T. designed the test setup and its printed circuit board. J.M.P, M.B., and D.Q. supervised the work. D.Q. wrote the initial version of the manuscript. All authors discussed the results and reviewed the manuscript.

\section*{Competing interests}
The authors declare no competing interests.

\section*{Data availability}
The data measured in this study are available from the corresponding author upon request.

\section*{Code availability} 
The software programs developed for this work are available online: \url{https://github.com/INTEGNANO/ForwardLearningUnderOneVolt}.

%%%%%%%%%%%%%%%%%%%%%%%%%%%%%%%%%%%%%%%%%%%%%%%%%%%%%%%%%%%%%%%%%%%%%%%%%%%%%%%%%%%%%%%%%%%%%%%%%%%%%

\section*{Methods}

\subsection*{Design and fabrication of hybrid memristor/CMOS circuits}

The hybrid circuits were produced in three stages, building on prior integration work  \cite{harabi2023memristor, jebali2024powering, bonnet2023bringing}. In the first stage, the complementary metal-oxide-semiconductor (CMOS) front end was manufactured at a commercial foundry using a 130-nm low-power technology node with four metal interconnect layers. The second stage involved the integration of hafnium oxide HfO$_x$ memristive devices on top of the fourth metal layer. Each device exhibits a stacked configuration consisting of titanium nitride (TiN), HfO$_x$, titanium (Ti), and a top TiN electrode. The HfO$_x$ functional layer was deposited by atomic layer deposition (ALD). Device patterning defined a circular geometry with a 300-nm diameter. Following this integration, a fifth metal interconnect layer was deposited above the memristor structures. In the final stage, the processed wafers were packaged into J-leaded Ceramic Chip Carrier (JLCC) modules by a commercial vendor. 

Two memristor platforms were used:
\begin{itemize}
    \item \textbf{MAC array (Figs.~\ref{fig:backprop1}b-d).} This integrated circuit targets in-memory multiply-accumulate (MAC) operations. It contains a 32$\times$64 two-transistor-one-resistor (2T1R) array; each cell integrates one memristor and two access transistors that enable (i) per-device forming, set, reset, and read via “vertical” selection and (ii) column-wise MAC via “horizontal” selection. Peripheral circuits provide the required biasing and current readout for column accumulation. Because synaptic weights are encoded as differential conductance pairs, the 2,048 devices implement up to 1,024 synapses. 

    \item \textbf{Large Array (Fig.~\ref{fig:backprop2}b).} This integrated circuit is a conventional 128$\times$64 one-transistor-one-resistor (1T1R) array with source lines shared by adjacent cells. Its periphery supports both digital and analog access \cite{harabi2023multimode}; in this work we use only direct analog device access for reading and programming. With differential pairing, the 8,192 devices implement up to 4,096 synapses.
\end{itemize}

The two platforms were fabricated on different wafers. The MAC array uses a newer generation of the HfO$_x$ integration flow optimized for lower-voltage programming. 
\changed{No device cherry-picking or fault masking was used in the learning demonstrations. For each network architecture, software synapses were assigned to physical differential pairs, in order, using a fixed deterministic address map before training. }

\subsection*{Sub-1~V reset regime characterization}

We first characterized the sub-1~V reset regime on the Large Array (Fig.~\ref{fig:weakreset}). Supplementary Note~1 details the rationale behind our forming strategy. Devices were  formed using a  constant-amplitude scheme: identical pulses were applied repetitively (one to five shots) until conduction appeared, rather than using a voltage ramp. To reinforce a thick, stable filament, three wakeup reset-set pairs were then applied. We subsequently recorded conductance trajectories under long sequences of low-amplitude reset pulses (5,000 shots). 

After a coarse sweep of pulse amplitude and width, we acquired fine-grained trajectories by applying 5,000   reset pulses at 0.9~V and 600~ns duration on a cohort of 1,268 devices. For each device, we quantified monotonicity/linearity of conductance change versus pulse count using a Pearson correlation coefficient between $G_i$ (post-$i$th-pulse conductance) and the pulse index $i$ over the first $P_{\max}$ pulses:
\begin{equation}
    \rho_G = \frac1{P_\text{max}}\sum_{i=1}^{P_{max}}\frac{(G_i-\mu_G)(i-(P_{max}+1) / 2)}{\sigma_G \sigma_P},
    \label{eq:pearson_coef}
\end{equation}
where $\mu_G$ and $\sigma_G$ are the mean and standard deviation over the considered pulse range, and $\sigma_P$ is a normalizing constant: $\sqrt{\sum_i(i - (P_\text{max}+1)/2)^2/P_\text{max}}$. Values of the Pearson coefficient near $-1$ indicate a nearly linear, strictly decreasing conductance with pulse number. The distribution of Pearson coefficients as well as sample conductance trajectories are presented in Fig.~\ref{fig:weakreset}e. 

 We also assessed endurance by repeating 300 cycles of 5,000 reset pulses at 0.9~V on four devices (Fig.~\ref{fig:weakreset}c). At the end of each of the 300 cycles, the devices were reinitialized using a high voltage reset and a set operation.

To assess retention of states created by sub-1~V reset, we initialized devices to a low-resistance state and then used only low-amplitude reset pulses to program 3,456 devices to conductances spanning 16-100~$\mu$S. Devices were stored under ambient laboratory conditions (23$^\circ$C) with no bias applied between measurements. Conductance was read at $t=$ 8, 20, and 90 days using the same read conditions employed elsewhere. Drift was defined as $\Delta G(t)=G(t)-G(0)$, and we report the distribution of $|\Delta G|$ across the population in Fig.~\ref{fig:weakreset}d.

We re-optimized the programming conditions to reproduce comparable sub-1~V reset behavior on the MAC array, which uses a newer, lower-voltage device flow. On this platform we employed a single wake-up (reset-set reinforcement) cycle instead of three during forming, and used 0.62~V, 30~ns reset pulses for low-voltage updates. All read/program conditions for both platforms are listed in Supplementary Note~2.
Note that all ``sub-1~V'' statements in this work refer to the voltage across the memristor terminals ($V_\text{BL}-V_\text{SL}$). To fully turn on the access transistors in the 130-nm arrays and reduce series resistance during reads/programs, the word lines are driven with a boosted gate voltage $>\!1$~V. This word-line overdrive is confined to the transistor gate; the memristor terminal voltage during reset-only updates remains below 1~V. This separation improves read accuracy and programming reproducibility without altering the device stress conditions, and would not be necessary in an advanced CMOS node.

\subsection*{Curation of the animal dataset}

We constructed a four-class dataset (brown bear, sloth bear, polar bear, giant panda) from a subset of ImageNet-1k images \cite{russakovsky2015imagenet}. After shuffling, images were split in a class-balanced manner into train (60\%, $\sim$800 per class), validation (30\%, $\sim$370 per class), and test (10\%, $\sim$135 per class) sets. The training split was augmented to 1,000 images per class using standard transforms (random horizontal flip, color jitter, and small rotations). All images were preprocessed to match the backbone (ResNet-18) input requirements (resize/crop and normalization consistent with ImageNet pretraining). Supplementary Fig.~4 shows sample images of the dataset.

For all simulations and hardware experiments, we removed the backbone’s final classifier and extracted the 512-dimensional penultimate feature vector for each image. %To emulate resource-constrained edge settings and reduce I/O, as well as make the classification task more challenging, 
We then restricted the feature space to 32 dimensions by retaining the first 32 channels of the 512-D vector (the same fixed subset for train/validation/test). 
\changed{This truncation serves two purposes in the present demonstration. First, it matches the physical constraints of the present hardware demonstration, allowing the whole trainable part to fit on the Large Array. Second, the truncation increases the difficulty of the task by making it less linearly separable. This effect is seen in Supplementary Figure~5, which compares the t-SNE between the full 512-dimensional representation and the restricted 32-dimensional representation. We see stronger class overlap after truncation.  }

The resulting 32-D features were cached and used identically across software baselines and on-chip experiments. 

\subsection*{Experimental Setup}
Both integrated circuits were mounted on in-house printed circuit boards (PCBs) and driven by an STM32F746ZG microcontroller, which generated array-specific digital control (addressing, mode selection, timing). Read/write biasing and electrical measurements were provided by a Keysight B1530A Waveform Generator/Fast Measurement Unit (WGFMU). A host computer coordinated the MCU and WGFMU; experiments were scripted in Python (PyTorch) with a lightweight wrapper that abstracts the matrix-vector multiplications between the operation on the MAC array and the emulation for the Large Array. 

\subsubsection*{Learning experiments with the MAC array (Fig.~\ref{fig:backprop1})}

The MAC array supports on-chip analog MAC and was  used for the perceptron experiments (32 inputs, 4 outputs) of Fig.~\ref{fig:backprop1}. To match the array’s three-level input interface, features were ternarized to $x_j\!\in\!\{-1,0,+1\}$. Simulations confirmed no measurable accuracy loss versus floating-point inputs for the bear-classification task, owing to the redundancy of the feature extractor.

\paragraph{Biasing.}
A differential weight is encoded as $w_{ij}\propto G^{+}_{ij}-G^{-}_{ij}$ using a pair of devices on adjacent columns. During inference, columns associated with $G^{+}$ and $G^{-}$ receive $+x_j$ and $-x_j$, respectively. Negative input levels are synthesized by offsetting the source line (sense node) to $V^{0}\!=\!0.7$\,V and driving the bit lines at
$V^{-}\!=\!0.5$\,V, $V^{0}$, or $V^{+}\!=\!0.9$\,V for input levels $x_j \in \{-1,0,+1\}$, yielding effective input voltages $V^{\mathrm{in}}_j \in\{-0.2,0,+0.2\}$\,V at the device. $V^{+}$ is supplied by a DC source, while $V^{-}$ and $V^{0}$ are generated by the WGFMU. 

\changed{Before a learning process, all devices are programmed with a Set operation. Thus, the initial neural-network weights are centered close to zero and their dispersion arises from device-to-device mismatch after this common electrical initialization. No device selection, remapping, or Write\&Verify initialization of random weights is used. A study on the impact of the initial state on the learning process is presented in Suppl. Note~3 (section ``Impact of weight initialization'').}

\paragraph{On-chip MAC and readout.}
Let $I_i$ be the column-summed current on output $i$. In differential encoding, columns for $G^{+}_{ij}$ and $G^{-}_{ij}$ are driven by $\pm x_jV_{read}$. The column current for class $i$ is therefore
\[
I_i = \sum_{j} \left(G^{+}_{ij}\,x_jV_{read} + G^{-}_{ij}\,(-x_j)V_{read}\right)
\;=\; \sum_{j} (G^{+}_{ij}-G^{-}_{ij})\,x_jV_{read}.
\]
A scalar gain $\kappa$ (software hyperparameter) maps $I_i$ to the perceptron logits $y_i=\kappa I_i$.

\paragraph{Row sequencing and software wrapper.}
Because the WGFMU channel count limits simultaneous drive/sense, MAC is executed row-by-row. A custom PyTorch module wraps the hardware calls and presents a \texttt{torch.nn.Linear}-like interface, configuring inputs and sequentially accumulating the on-chip MAC to form the output tensor. Training/validation/test metrics are thus computed with the same on-chip MAC path.

\paragraph{Training loop.}
Each update step consists of: (i) hardware forward pass on a mini-batch to obtain $\mathbf{y}$; (ii) gradient computation; % using Eq.~\eqref{eq:grad_bp}; 
(iii) thresholding small gradients to zero (see Main text); and (iv) a programming sweep applying a single sub-1~V reset pulse to the appropriate device of each differential pair (pulse on $G^{+}$ for a positive gradient, i.e. a negative update, pulse on $G^{-}$ for a negative gradient, i.e. a positive update). A subsequent read sweep acquires $G^{+},G^{-}$ for logging and diagnostics.

\subsubsection*{Learning experiments with the Large Array (Figs.~\ref{fig:backprop2} and \ref{fig:forward})}

Because the MAC array is limited to 32$\times$64 devices, multilayer backpropagation and all forward-only experiments were performed on the Large Array (128$\times$64). This platform does not implement on-chip MAC; instead, MAC operations are emulated in software while all weight reads/writes are executed on hardware. The equivalence of on-chip versus software MAC was verified on the perceptron using the MAC array (see Fig.~\ref{fig:backprop1}f).

\paragraph{Weight mapping.} After each hardware read, software weights are reconstructed from device conductances via a fixed scale factor $s$:
\begin{equation*}
W \;=\; s\,\left(G^{+}-G^{-}\right),   
\end{equation*}
with $(G^{+},G^{-})$ the differential pair assigned to each synapse. A static address map ties each software weight to its physical cell pair. This scale factor $s$ is related to the scalar gain $\kappa$ introduced during on-chip MAC operations to map the read current to the perceptron logits: $s = \kappa V_\mathrm{read}$.

\changed{As in the case of the MAC array, before a learning process, all devices are programmed with a Set operation.}

\paragraph{Training loop (hardware-in-the-loop).} Following an initial array read, each update step proceeds as:
\begin{enumerate}\itemsep0pt
\item Software inference using the current $W$ to obtain activations/logits.
\item Gradient computation for the chosen loss: backpropagation, SFF (eq.~\eqref{eq:sfffinal}), or competitive forward (eq.~\eqref{eq:cffinal}).
\item Gradient sparsification: entries below the threshold $\tau$ are zeroed; for layerwise schedules, frozen layers are masked.
\item Programming sweep: for each selected synapse, apply a single sub-1\,V reset pulse to the appropriate device of the differential pair (pulse on $G^{+}$ for a positive gradient, i.e. a negative update, pulse on $G^{-}$ for a negative gradient, i.e. a positive update). 
\item Verification read: re-read $(G^{+},G^{-})$ to update $W$ and log statistics.
\end{enumerate}

Backpropagation models are trained layerwise from output to input (Fig.~\ref{fig:backprop2}); forward-only models are trained layerwise from input to output using their local losses (Fig.~\ref{fig:forward}).

\changed{A complete physical-array bear-classification training run (using competitive Forward) takes 6.5~h in the present laboratory setup, whereas the corresponding device-calibrated simulation takes 11~s on an NVIDIA GeForce GTX 1080 Ti. The experimental wall-clock time is dominated by serial laboratory instrumentation and host-instrument communication, not by the intrinsic duration of the memristor read and reset pulses.}

\changed{\subsection*{Read-disturb and Write\&Verify trained-model comparison}}

\changed{
For the read-disturb experiment, we first trained a competitive forward network on the bear classification task using the sub-1~V reset-only update rule. After training, and without applying additional programming pulses, all mapped memristors were read every 2~s for 25~h using the standard Large Array read conditions. The network weights were reconstructed from the measured differential conductances, and the test accuracy was evaluated every 200~s.}

\changed{To construct a matched conventional programming baseline, we used the final weights of the same competitive forward training run. The weights were quantized to six conductance levels. In software, this quantized model reached 89.71\% test accuracy, compared with 90.62\% before quantization. The quantized weights were then programmed on the physical array using a set-based three-trial Write\&Verify procedure. The test accuracy immediately after programming was 90.07\%. The same read-disturb protocol was then applied to this Write\&Verify-programmed model.}

\changed{In Fig.~\ref{fig:practical_stability}a, the reported curves are smoothed using a moving window of 50 accuracy values. The solid line is the mean over the window and the shaded region is the corresponding standard deviation.}

\changed{\subsection*{Repeated task reprogramming}}

\changed{To test whether the same physical array can be reprogrammed for different inference tasks, we alternated hardware training between the bear and bird four-class transfer-learning tasks. Both tasks used the same ResNet-18 feature-extraction procedure, the same competitive forward architecture, and identical hyperparameters. Each task was trained for 30 epochs. The last layer was trained only during the final 10 epochs. Between consecutive tasks, all memristors were globally reinitialized by hard RESET followed by SET. No device selection, remapping, or task-specific hyperparameter optimization was performed between task switches.}

\subsection*{Perceptron and layerwise backpropagation-based learning algorithm}

We train either a single fully connected layer (perceptron, Fig.~\ref{fig:backprop1}) or a two-layer multilayer perceptron (MLP,  Fig.~\ref{fig:backprop2}), both without biases. The two-layer model uses 48 hidden units with ReLU activation; both architectures use a softmax output and cross-entropy loss. No normalization is applied to inputs or activations, except in the on-chip MAC experiments where inputs are ternarized to $x_j\in\{-1,0,+1\}$ to match the hardware interface. 

The perceptron is trained for 20 epochs. The two-layer MLP is trained for 30 epochs in a layerwise fashion: the output layer is trained for the first 10 epochs, then frozen while the input (hidden) layer is trained for the remaining 20 epochs. 

We use minibatches of size 16. A batch-size study is presented in Supplementary Note~3 \changed{(section ``Simulation study of the effect of batch size'')}. When MAC is performed on-chip, gradients are computed analytically from the measured outputs;
%using Eq.~\eqref{eq:grad_bp}; 
when MAC is emulated, PyTorch autograd is used.

\subsection*{Supervised Forward-Forward learning rule}
We summarize here  the specific choices used for implementing Supervised Forward-Forward learning in the experiments. For each training example with label $y\in\{1,\dots,C\}$, we form positive inputs by concatenating the feature vector with the corresponding one-hot label token, and negative inputs by concatenating the same feature with a randomly chosen incorrect label token. For layer $\ell$, let $h_\ell^+$ and $h_\ell^-$ denote the activations produced by the positive and negative inputs, respectively, and let the layer ``goodness'' be $g_\ell(h)=\eta_\ell \|h_\ell\|_2^2$. All SFF experiments use $\eta=1$.

The per-layer loss combines positive and negative goodness terms:
\begin{equation}
    \mathcal{L}_{\ell}^{SFF} = -\frac{1}{2} \left[ \log \sigma \left(g_\ell(h_\ell^+) - \eta_\ell\theta_\ell^+N_h\right) + \log \left(1 - \sigma(g_\ell(h_\ell^-)-\eta_\ell\theta_\ell^-N_h)\right) \right]
    \label{eq:sff_loss}
\end{equation}
where $\sigma$ is the logistic sigmoid, $N_h$ is the number of neurons on the output side of layer $\ell$ (i.e. the number of activations) and the $\theta_\ell^{\pm}\in\mathbb{R}$ are determined through hyperparameter optimization. Given a minibatch $\mathcal{B}$ of $N_\mathcal{B}$ positive/negative activations, the gradient with respect to the weight matrix $W_\ell$ of layer $\ell$ is
\begin{equation}
    \nabla_{W_\ell} \mathcal{L}_{\ell}^{SFF} = \frac1{2N_\mathcal{B}} \sum_{{h_\ell^+, \ h_\ell^-} \in \mathcal{B}}\left[-\frac{h_\ell^+ \nabla_{W_\ell} (h_\ell^+)}{1+e^{g_\ell(h_\ell^+) - \eta_\ell\theta_\ell^+N_h}} + \frac{h_\ell^- \nabla_{W_\ell} (h_\ell^-)}{1+e^{\eta_\ell\theta_\ell^-N_h - g_\ell(h_\ell^-)}}\right].
    \label{eq:grad_sff}
\end{equation}

Considering the Rectified Linear Unit (ReLU) as the activation function, the above gradient with respect to the weight parameter $w_{i,j}$ in layer $\ell$ can be expressed  more simply as
\begin{equation}
    \nabla_{i,j} = -\sum_{n=1}^{N_\mathcal{B}} \left[
    \frac{h_{n,i}^{+} x_{n,j}^{+}}{D^+_n}
    -
    \frac{h_{n,i}^{-} x_{n,j}^{-}}{D^-_n}
    \right].
    \label{eq:sfffinal}
\end{equation}
where $D^+_n = {2N_\mathcal{B}}\left[1 + \exp\left(\,g(h^{+}_n) - \eta \theta^{+}N_h\right)\right]$ and $D^-_n = {2N_\mathcal{B}}\left[1 + \exp\left(\,\eta \theta^{-}N_h - g(h_n^{-})\right)\right]$.

This learning rule is local: the weight update for synapse connecting two neurons depends only on the activities of the neurons to which it is connected. To implement it in hardware one only needs to store the $D^\pm_n$, $h_{n,i}^{\pm}$, and $x_{n,j}^{\pm}$ over the minibatch and then compute eq.~\eqref{eq:sfffinal}. The memory cost is of $N_\mathcal{B}(2+2N_x+2N_h)$ parameters, where $N_x$ and $N_h$ are the number of neurons on the input and output side of the layer. This cost grows linearly with the number of neurons $N_x+N_h$ (and not number of synapses $N_xN_h$), therefore the requirement in working memory is sustainable (see Discussion).

Usually, an SFF layer is followed by layer normalization, to eliminate the goodness and promote the learning of new features by subsequent SFF layers\cite{hinton2022forward}. Here, layer normalization is not needed as our experiments use a single SFF layer followed by a cluster-based, competitive forward head, which is trained only on the positive examples (see next section).
 
\subsection*{Learning rule used for the \changed{readout} head of SFF experiments and for competitive forward learning}
We now detail the learning rule used in our experiments for the second layer (reading head) of SFF experiments and for both layers for competitive forward learning. Following \cite{Papachrist24ClusterF}, each layer $\ell$ partitions its activations into class-associated clusters. 
As in SFF, layer goodness is defined as $g_\ell(h)=\eta_\ell \|h_\ell\|_2^2$.
When using $\eta=1$, for an example with label $y$, the activity of the cluster tied to $y$ is encouraged to increase, while the activity of all other clusters is discouraged. This requires only a single forward pass (no label injection into the input).

In our competitive forward experiments, we use $\eta=1$ for the final layer, and $\eta=-1$ for the first layer. When using $\eta=-1$, for an example with label $y$, the activity of all clusters not tied to $y$ is encouraged to increase, while the activity of  the cluster tied to $y$ is discouraged. \changed{Using $\eta=-1$ for the first layer is a task-dependent heuristic leading to a modest improvement in the accuracy-pulse trade-off in our two-layer learning experiments (see Suppl. Fig.~7), and is not a universal prescription for deeper network.}

Let $h_\ell$ denote the activations at layer $\ell$, $Z_\ell$ a binary mask selecting the cluster corresponding to the true class (and $(1-Z_\ell)$ the complement), $C$ the number of classes, and $g_\ell(h)$ the layer “goodness” . The per-layer loss is
\begin{equation}
    \mathcal{L}_{\ell}^{CF} = -\frac{1}{2} \left[ \log \sigma \left(g_\ell(h_\ell Z_\ell)-\eta_\ell\theta_\ell^+\right) + \log \left(1 - \sigma\left(g_\ell(h_\ell(1-Z_\ell))-\eta_\ell\theta_\ell^-\right)\right) \right],
    \label{eq:scf_loss}
\end{equation}
where $\sigma$ is the logistic sigmoid and $\theta_\ell^{\pm}\in\mathbb{R}$ are offsets.

In hardware we found it beneficial to use a temperature-based variant in which $\theta_\ell^{\pm}$ scale the sigmoid arguments (rather than shift them):
\begin{equation}
    \mathcal{L}_{\ell}^{CF} = -\frac{1}{2} \left[ \log \sigma \left(\theta_\ell^+g_\ell(h_\ell Z_\ell)\right) + \log \left(1 - \sigma\left(\theta_\ell^-g_\ell(h_\ell(1-Z_\ell))\right)\right) \right].
    \label{eq:scft_loss}
\end{equation}
Given a minibatch $\mathcal{B}$ of $N_\mathcal{B}$ layer inputs, the corresponding gradient with respect to the weight matrix $W_\ell$ of layer $\ell$ is
\begin{equation}
    \nabla_{W_\ell} \mathcal{L}_{\ell}^{CF} =\frac1{2N_\mathcal{B}} \sum_{{h_\ell} \in \mathcal{B}}\left[-\frac{\theta_\ell^+ (h_\ell \odot Z_\ell) \nabla_{W_\ell} (h_\ell \odot Z_\ell)}{1+e^{\theta_\ell^+ g_\ell(h_\ell\odot Z_\ell)}} +
    \frac{\theta_\ell^- (h_\ell \odot(1-Z_\ell)) \nabla_{W_\ell} (h_\ell \odot (1-Z_\ell))}{1+e^{- \theta_\ell^-g_\ell(h_\ell\odot(1-Z_\ell))}}\right],
    \label{eq:grad_cf_temp}
\end{equation}
where $\odot$ denotes element-wise multiplication with the binary masks. This formulation preserves fast convergence and allows separate control of the positive/negative contributions via $\theta_\ell^{\pm}$.
Considering the Rectified Linear Unit (ReLU) as the activation function, the above gradient of the loss with respect to the weight parameter $w_{i,j}$ in layer $\ell$ can be expressed more simply as
\begin{equation}
    \nabla_{i,j} = -\sum_{n=1}^{N_\mathcal{B}}  h_{n,i} x_{n,j}
    \left[
    \frac{  \delta \left(C(i)=Y_n \right)  }{D^+_n}   
    -
    \frac{\delta \left(C(i) \neq Y_n \right)}{D^-_n} 
    \right],
    \label{eq:cffinal}
\end{equation}
where $D^+_n = 2N_\mathcal{B}\cdot\left[1+\exp\left(\,\theta^+ g(h_n \odot Z_n)\right)\right]\textfractionsolidus\theta^+$ and $D^-_n = 2N_\mathcal{B}\cdot\left[1+\exp\left(\,-\theta^-g(h_n\odot(1-Z_n))\right)\right]\textfractionsolidus\theta^-$.

%$D^-_n = (C-1)\left[1+\exp\left(\,-\frac{\theta^-}{C-1}g(h_n\odot(1-Z_n))\right)\right]/\left[2N_\mathcal{B}\cdot\theta^-\right]$

$C(i)$ is the class number corresponding to the cluster of neuron $i$. $\delta(C(i)=Y_n)$ is one when this class corresponds to the true class $Y_n$, zero otherwise (and vice-versa for $\delta(C(i) \neq Y_n)$).

As for SFF, this learning rule is local: the weight update for synapse connecting two neurons depends only on the activities of the neurons to which it is connected. 
To implement it in hardware one only needs to store the $D^\pm_n$, $Y_n$, $h_{n,i}$, and $x_{n,j}$ over the minibatch and then compute eq.~\eqref{eq:cffinal}. The memory cost is $N_\mathcal{B}(3+N_x+N_h)$, where $N_x$ and $N_h$ are the number of neurons on the input and output side of the layer. This number grows linearly with number of neurons $N_x+N_h$ (and not number of synapses $N_xN_h$), therefore the requirement in working memory is sustainable (see Discussion), and lower than for SFF.

\subsection*{Statistical analysis of the experimental results}
The final test accuracies for the two-layer perceptron trained with backpropagation are 90.62\%, 91.18\%, 89.89\%, 87.87\%, 90.44\%. For SFF, the accuracies are 88.05\%, 90.44\%, 87.68\%, 89.89\%, 91.36\%. For competitive forward, the accuracies are 91.18\%, 90.62\%, 89.52\%, 90.44\%, 86.03\%. 
We tested pairwise null hypotheses of equal means: $H_0^\mathrm{bp-sff} : \mu_\mathrm{bp} = \mu_\mathrm{sff}$, $H_0^\mathrm{bp-cf} : \mu_\mathrm{bp} = \mu_\mathrm{cf}$, $H_0^\mathrm{sff-cf} : \mu_\mathrm{sff} = \mu_\mathrm{cf}$; where $\mu_x$ is the averaged final accuracies for the learning rule $x$.
By computing a two-tailed Welch's t-test, we get $p^\mathrm{bp-sff}=0.586$, $p^\mathrm{bp-cf}=0.697$, $p^\mathrm{sff-cf}=0.951$. All p-values largely exceed $\alpha=0.05$. The conclusion is unchanged after Holm-Bonferroni correction. 

\subsection*{Energy consumption analysis of the learning process}

During training we logged, for every reset event, the device address and the conductance $G$ measured immediately before. The energy delivered to a memristor during a single reset pulse was estimated  as $E_\mathrm{pulse}=GV^2_\mathrm{reset}t_\mathrm{reset}$, where $V_\mathrm{reset}$ and $t_\mathrm{reset}$ are the pulse voltage and duration used for sub-1~V reset (see Suppl. Note~2). The total programming energy for a run is the sum of \(E_\mathrm{pulse}\) over all pulses applied to the active device of each differential pair. 

To project the programming energy with the lower-voltage (MAC-array) device technology, we recomputed $E_\mathrm{pulse}$ for the exact sequence of pulses recorded on the Large Array but substituting the MAC-array pulse parameters (Suppl. Note~2). For the program-and-verify baseline, we used its prescribed program pulses and verify reads (amplitudes, widths) from a prior work using hafnium oxide memristors\cite{esmanhotto2022experimental}; the reported baseline energy includes the mean number of attempts and verify reads until the target criterion is met reported in this work.

For array reads (vector-matrix products), energy is dominated in our experimental setup by the read pulse due to the relatively long integration time. We therefore estimate the read energy by summing the read pulse energy over all forward passes that need to be executed during training (one pass per layer for competitive forward; two passes for SFF).
By contrast, in a fully integrated CMOS implementation with short read pulses, peripheral circuits (e.g., ADCs) dominate energy. To estimate a realistic end-to-end MAC cost (including ADC), we use the measured energy efficiency of a state-of-the-art foundry in-memory-computing  platform in 22\,nm  using memristors\cite{xue2021cmos}. For the precision closest to our use case (4-bit weights, 2-bit input activations, 10-bit outputs), the reported efficiency in\cite{xue2021cmos}  is \(57.5~\mathrm{TOPS/W}\). We convert the total number of MAC operations executed during training into an energy based on this reported efficiency (note that in this context, one MAC is counted as two operations: one multiply plus one add).

\subsection*{Computer setup}
All hardware experiments were orchestrated from a workstation equipped with an Intel Xeon E5-2640 CPU. Control software was written in Python 3.11 using the PyTorch 2.7 framework and handled coordination of the WGFMU and microcontroller, dataset handling, and training loops.  All simulations reported in the Supplementary Notes were executed on a server with an NVIDIA GeForce RTX~2080~Ti (11~GB VRAM). These simulations were implemented in Python 3.9 using the JAX and JAXLIB 0.4 framework.

\bibliography{references}  

%TC:endignore

\end{document}